# Performance evaluation of energy detector over generalized non-linear and shadowed composite fading channels using a Mixture Gamma Distribution

He Huang, et. al
Beijing University of Posts and elecommunications, Beijing 100876, China

The performance of energy detection (ED) for independent and identically distribution (i.i.d.) signal models is analyzed over generalized composite non-linear line-of-sight (LOS) and non-line-of-sight (NLOS) shadowed fading scenarios. The novel expressions for $\alpha$-$\kappa$-$\mu$/Gamma and $\alpha$-$\eta$-$\mu$/Gamma fading channels have been derived to approximate by using the mixture gamma (MG) distribution under low instantaneous signal-to-noise (SNR) condition. On the basis of the deduced fading distributions, novel, exact and close-form detective models are derived to evaluate the sensing performance with different key fading parameters over generalized non-linear and shadowed composite fading channels.

*Introduction:* Energy detection (ED) is a low-complexity valid spectrum sensing (SS) technology which is widely applied in unknown deterministic non-coherent signals processing in cognitive radios (CRs) [1], and it has been extensively used for the performance analysis of complicated and extreme multipath fading channels due to the advantage of detection without a prior knowledge and implementation simplicity [2-5]. Numerous types of detectors have been employed to detect primary signals in multiple communication systems (such as multicarrier system with orthogonal frequency-division multiplexing, multi-antenna communication and covariance-based detection etc) [6].

It is also known that the investigation of the effective performance evaluation of general short-term multipath physical fading models (for example, $\alpha$-$\mu$, $\kappa$-$\mu$ and $\eta$-$\mu$ etc) have been regarded as a critical work for unified small-scale models include classical cases as Rayleigh, Nakagami-m, Hoyt, Weibull and Rice distributions [7]. Furthermore, non-linear propagation medium models have been proposed to solve the problem of limited accuracy and generality for the patterns listed above [3, 4]. On the other hand, the composite fading distribution which combines short term and long term (e.g., gamma distribution) is applied to describe the intertwined effects of short-term and long-term fading in wireless communication [6-8]. To sum up, it is extremely meaningful to derive unified detective models over non-linear and shadowed composite fading channels for various ED models.

In this letter, we derive the novel exact expressions for the average detection probability of general ED models over $\alpha$-$\kappa$-$\mu$/Gamma and $\alpha$-$\eta$-$\mu$/Gamma fading channels with mixture gamma (MG) distribution. To the best of the present authors` work, these types of channels with MG distribution have not been studied for i.i.d. ED signals samples in the open literature. Besides, exact rational expressions with non-integer fading variables are deduced to solve the problem of sensing optimization for closed-form evaluation of generalized composite fading channels.

*ED model:* The system block diagram of general model of ED is shown in Fig.1, the primary signal follows a binary hypothesis: $H_0$ ($y(n)=w(n)$, signal absent), $H_1$ ($y(n)=hs(n)+w(n)$, signal present). Where $n=1,2,...,N$, $N$ is the number of simples, $y(n)$ is the received signal, $h$ is the wireless channel gain, $w(n)$ is assumed i.i.d. circularly symmetric complex Gaussian (CSCG) noise which obeys $N(0, \sigma_w^2)$, $s(n)$ is the primary signal which obeys $N(0, \sigma_s^2)$, $\lambda$ is the threshold of ED, $\gamma$ is the instantaneous SNR and $\gamma=(|h|\sigma_s/\sigma_w)^2$ [6].

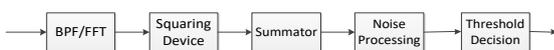

**Fig. 1** *general model of energy detection*

In Fig.1 the binary hypothesis can be derived as Eq. (1), if $s(n)$ is CSCG random variable, $\xi(\gamma)=2\gamma+\gamma^2$, else if $s(n)$ is complex valued phase shift keying (PSK), $\xi(\gamma)=2\gamma$.

$$H_0 \sim N(N\sigma_w^2, N\sigma_w^4); H_1 \sim N(N\sigma_w^2(1+\gamma), N\sigma_w^4(1+\xi(\gamma))) \quad (1)$$

From (1) the probability of false alarm and detection can be expressed as at low signal-to-noise (SNR) respectively, for $\gamma$ has little impact on the variance under $H_1$.

$$P_f = \frac{1}{2}erfc(\frac{\lambda - N\sigma_w^2}{\sqrt{2N\sigma_w^2}}) \quad (2)$$

$$P_d = \frac{1}{2}erfc(\frac{\lambda - N\sigma_w^2(1+\gamma)}{\sqrt{2N\sigma_w^2(1+\xi(\gamma))}}) \approx \frac{1}{2}erfc(\frac{\lambda - N\sigma_w^2(1+\gamma)}{\sqrt{2N\sigma_w^2}}) \quad (3)$$

*Average detection probability over composite shadowed fading channels:* The probability density function (PDF) of composite shadowed fading channels can be expressed as [7, 8]

$$f_\gamma^{channel/Gamma}(\gamma) = \int_0^\infty f_{\gamma|Y}^{channel}(\gamma|y) f_Y^{Gamma}(y) dy \quad (4)$$

where $f_Y^{Gamma}(y) = y^{k-1}e^{-y/\Omega}/(\Gamma(k)\Omega^k)$, $k$ is the shaping parameter, $\Omega$ is the average SNR and $\Gamma(.)$ is the gamma function.

With (4), [3, Eq. (8)] and [5, Eq. (1)], the PDF of non-linear shadowed composite fading channels can be expressed as

$$f_\gamma^{\alpha-\kappa-\mu/Gamma}(\gamma) = \frac{\alpha\mu\kappa^{\frac{1-\mu}{2}}(1+\kappa)^{\frac{1+\mu}{2}}}{2e^{\kappa\mu}} \int_0^\infty \frac{\gamma^{\frac{\alpha(1+\mu)}{4}-1}}{y^{\frac{\alpha(1+\mu)}{4}}}\exp(-\mu\frac{\gamma^{\frac{\alpha}{2}}}{y^{\frac{\alpha}{2}}} - \kappa\mu\frac{\gamma^{\frac{\alpha}{2}}}{y^{\frac{\alpha}{2}}}) \cdot$$
$$I_{\mu-1}(2\mu\sqrt{\kappa(1+\kappa)}\frac{\gamma^{\frac{\alpha}{4}}}{y^{\frac{\alpha}{4}}}) \frac{y^{k-1}e^{-\frac{y}{\Omega}}}{\Gamma(k)\Omega^k} dy \quad (5)$$

$$f_\gamma^{\alpha-\eta-\mu/Gamma}(\gamma) = \frac{\sqrt{\pi}\alpha h^\mu \mu^{\mu+\frac{1}{2}}}{\Gamma(\mu)H^{\mu-\frac{1}{2}}} \int_0^\infty \frac{\gamma^{\frac{\alpha}{2}(\mu+\frac{1}{2})-1}}{y^{\frac{\alpha}{2}(\mu+\frac{1}{2})}} \exp(-\frac{2\mu h\gamma^{\frac{\alpha}{2}}}{y^{\frac{\alpha}{2}}}) I_{\mu-\frac{1}{2}}(\frac{2\mu H\gamma^{\frac{\alpha}{2}}}{y^{\frac{\alpha}{2}}}) \cdot$$
$$\frac{y^{k-1}e^{-\frac{y}{\Omega}}}{\Gamma(k)\Omega^k} dy \quad (6)$$

where $\alpha$ is the nonlinear characteristics of the propagation medium, $\kappa$ is the ratio between the total power of the dominant components and the total power of the scattered waves, $\mu$ represents the number of multipath waves. $\eta$ is the power ratio between the in-phase and quadrature components in format I, $h=(1+\eta)^2/(4\eta)$, $H=(1-\eta^2)/(4\eta)$ and $0<\eta<\infty$. In format II $\eta$ is the correlation coefficient between the in-phase and quadrature components, $h=1/(1-\eta^2)$, $H=\eta/(1-\eta^2)$ and $-1<\eta<1$. $I_o(.)$ denotes the modified Bessel function of the first kind with the order $o$.

From [6, Eq. (5)], the PDF of MG can be expressed as

$$f_\gamma^{MG}(\gamma) = \sum_{v=1}^S \tilde{\alpha}_v \gamma^{\beta_v-1} e^{-\zeta_v \gamma} \quad (7)$$

where $S$ is the number of terms, $\tilde{\alpha}_v$, $\beta_v$ and $\zeta_v$ are parameters that represent the potential fading and shadowing effects. The integration in (5) and (6) can be approximated with Gaussian-Laguerre quadrature sum as $\int_0^\infty e^{-x}\varsigma(x)dx = \sum_{i=1}^N \omega_i \varsigma(x_i)$, $x_i$ and $\omega_i$ are the abscissas and weight factors respectively [9, Eq. (25.4.45)].

Simplifying (5) and (6) with functional transformation for the form of (7) [10] like [7, Eq. (3)-Eq. (9)],

$$\varphi(\alpha,\kappa,\mu) = \frac{\mu^{\frac{2}{\alpha}k-\frac{\mu}{2}+\frac{1}{2}}(1+\kappa)^{\frac{2}{\alpha}k}}{\Gamma(k)\Omega^k \kappa^{\frac{\mu-1}{2}} e^{\kappa\mu}}, \theta_i = \varphi(\alpha,\kappa,\mu)\omega_i x_i^{\frac{2}{\alpha}k+\frac{\mu}{2}-\frac{1}{2}} I_{\mu-1}(2\sqrt{\kappa\mu x_i}) \quad (8)$$

$$f_\gamma^{MG:\alpha-\kappa-\mu/Gamma}(\gamma) = \sum_{i=1}^S \tilde{\alpha}_i \gamma^{\beta_i-1} e^{-\zeta_i \gamma}|_{(\tilde{\alpha}_i,\beta_i,\zeta_i)}, x = \frac{\gamma^{\frac{\alpha}{2}}(\kappa\mu+\mu)}{y^{\frac{\alpha}{2}}} \quad (9)$$

$$\tilde{\alpha}_i = \frac{\varphi(\alpha,\kappa,\mu)\theta_i}{\sum_{l=1}^S \theta_l \Gamma(\beta_l)\zeta_l^{-\beta_l}}, \beta_i = k, \zeta_i = \frac{(\mu+\kappa\mu)^{\frac{2}{\alpha}}}{\Omega x_i^{\frac{2}{\alpha}}}$$

$$\varphi(\alpha,\eta,\mu) = \frac{2^{\frac{2}{\alpha}k-\mu+\frac{1}{2}}\sqrt{\pi}h^\mu \mu^{\frac{2}{\alpha}k-\frac{1}{2}+\frac{2}{\alpha}}}{\Gamma(\mu)\Gamma(k)H^{\mu-\frac{1}{2}}\Omega^k}, \theta_m = \psi(\alpha,\eta,\mu)\omega_m x_m^{-\frac{2}{\alpha}k+\mu-\frac{1}{2}} I_{\mu-\frac{1}{2}}(\frac{Hx_i}{h}) \quad (10)$$

$$f_\gamma^{MG:\alpha-\eta-\mu/Gamma}(\gamma) = \sum_{m=1}^S \tilde{\alpha}_m \gamma^{\beta_m-1} e^{-\zeta_m x}|_{(\tilde{\alpha}_m,\beta_m,\zeta_m)}, x = \frac{2\mu h\gamma^{\frac{\alpha}{2}}}{y^{\frac{\alpha}{2}}}$$

$$\tilde{\alpha}_m = \frac{\psi(\alpha,\eta,\mu)\theta_m}{\sum_{l=1}^S \theta_l \Gamma(\beta_l)\zeta_l^{-\beta_l}}, \beta_m = k, \zeta_m = \frac{(2\mu h)^{\frac{2}{\alpha}}}{\Omega x_m^{\frac{2}{\alpha}}} \quad (11)$$

Further the average probability of detection can be expressed as

$$\bar{P}_d = \int_0^\infty P_d^{ED} f_\gamma^{MG}(\gamma) d\gamma \quad (12)$$

By using the identity $erfc(-x)+erfc(x)=2$, with (3), (9) and (11), (12) can be derived as

$$\bar{P}_d = \int_0^\infty (1 - \frac{1}{2}erfc(C_1\gamma + C_2)) \sum_{v=1}^S \tilde{\alpha}_v \gamma^{\beta_v-1} e^{-\zeta_v \gamma} d\gamma$$
$$= 1 - \frac{1}{2} \sum_{v=1}^S \tilde{\alpha}_v \int_0^\infty \gamma^{\beta_v-1} e^{-\zeta_v \gamma} erfc(C_1\gamma + C_2) d\gamma \Big|_{(C_1,C_2)=(\frac{\sqrt{N}}{\sqrt{2}}, \frac{\sqrt{N}}{\sqrt{2}}, \frac{\lambda}{\sqrt{2}\sqrt{N}\sigma_w^2})} \quad (13)$$

With the aid of [10, Eq. (2.8.9.1)], (9) and (11), (13) can be deduced as

$$\bar{P}_d = 1 - \frac{1}{2} \sum_{v=1}^S \tilde{\alpha}_v (-1)^{\beta_v-1} \frac{\partial^{(\beta_v-1)}}{\partial \zeta_v^{(\beta_v-1)}} (\frac{1}{\zeta_v}(erfc(C_2) - e^{\frac{\zeta_v^2 + 4\zeta_v C_1 C_2}{4C_1^2}} erfc(C_2 + \frac{\zeta_v}{2C_1}))) \quad (14)$$

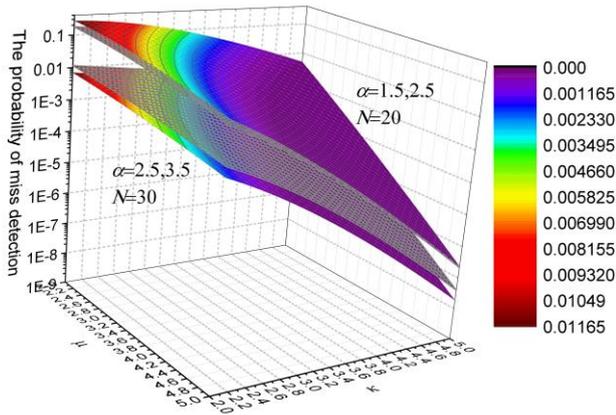

**Fig.2** *The probability of miss detection versus κ and μ for different values of α and N, k=4, Ω=-5dB, $\sigma_w^2$=1, $P_f$=0.01*

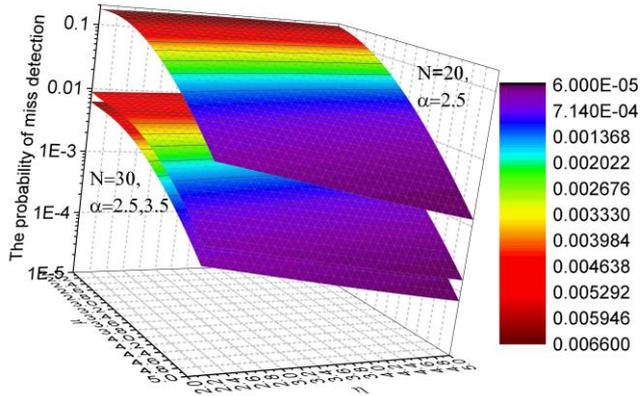

**Fig.3** *The probability of miss detection versus η and μ for different values of α and N in format* I *, k=4, Ω=-5dB, $\sigma_w^2$=1, $P_f$=0.01*

*Numerical results:* Fig.1 and 2 present the probability of miss detection versus κ, η and μ for different values of α at low SNR condition. It can be observed that the capacity of detection improves as α increases because higher nonlinear fading parameter is corresponding to more power value to capture the nonlinear propagation effects, in addition, the improvement for the power ratio between dominant components and scattered waves and the multipath waves number will also cut down the miss detection probability. Furthermore, higher value of multipath waves has a more important effect for detection performance. However, the increases of η will reduce the detection probability because it represents the power ratio between in-phase and quadrature components of multiple paths. On the other hand, more importantly, as the number of simples for general models of ED increase, the probability of miss detection can be obviously improved and the ED algorithm has remarkable performance on optimizing sensing capacity over generalized non-linear and shadowed composite fading channels.

*Conclusion:* In this letter, the performance of general models of ED under low SNR condition has been analyzed over generalized composite non-linear LOS and NLOS shadowed fading scenarios. Novel and exact close-form expressions for α-κ-μ/Gamma, α-η-μ/Gamma fading channels and the average detection probability with ED have been derived by using MG distribution to optimize the sensing performance. Numerical results demonstrate that the sensing performance can be effectively evaluated and optimized both for increasing ED simples and improving the corresponding fading parameters.